\documentclass[a4paper]{PoS}
\pdfoutput=1

\usepackage {amsmath}
\usepackage {amsfonts}
\usepackage{feynmp}
\usepackage{slashed}
\usepackage{subfig}
\usepackage{pstool}
\usepackage{comment}

\title{Naturalness of effective theories in Wilsonian approach}
\ShortTitle{Naturalness of effective theories in Wilsonian approach}
\author{\speaker{T. Krajewski}\\ 
University of Warsaw\\
E-mail: \email{Tomasz.Krajewski@fuw.edu.pl}} 
\author{Z. Lalak\\ 
University of Warsaw\\
E-mail: \email{Zygmunt.Lalak@fuw.edu.pl}} 
\abstract{
We have computed Wilsonian effective action in a simple model containing scalar field with quartic self-coupling which interacts via Yukawa coupling with a Dirac fermion. The model is invariant under a chiral parity operation, which can be spontaneously broken by a vev of the scalar field. We have computed explicitly Wilsonian running of relevant parameters which makes it possible to discuss in a consistent manner the issue of fine-tuning and stability of the scalar potential. This has been compared with the typical picture based on Gell-Mann--Low running. Since Wilsonian running includes automatically integration out of heavy degrees of freedom, the running differs markedly from the Gell-Mann--Low version. However, similar behavior can be observed: scalar mass squared parameter and the quartic coupling can change sign from a positive to a negative one due to running which causes spontaneous symmetry breaking or an instability in the renormalizable part of the potential for a given range of scales.
As for the issue of fine-tuning, since in the Wilsonian approach power-law terms are not subtracted, one can clearly observe the quadratic sensitivity of fine-tuning measure to the change of the cut-off scale. 
}
\FullConference{18th International Conference From the Planck Scale to the Electroweak Scale \\
		25-29 May 2015\\
		Ioannina, Greece }
\ShortTitle{Naturalness in Wilsonian approach}
\begin{document}
\maketitle
\section{Introduction}
The recent discovery of the Higgs boson at the Large Hadron Collider \cite{Aad:2012tfa,Chatrchyan:2012ufa}, promotes the question about the protection of the electroweak breaking scale to one of the most puzzling problems of fundamental physics. The observed compatibility of properties of the newly observed particle, with predictions coming from the Standard Model, additionally strengthens tension between the standard theoretical reasoning which results in a~prediction of new physics near the electroweak scale and reality. This situation strengthens the need of revisiting the naturalness principle. 

Numerous authors \cite{Aoki:2012xs,Farina:2013mla,Jegerlehner:2013cta,deGouvea:2014xba,Bar-Shalom:2014taa} propose new definitions of the naturalness. Our goal is less ambitious. We shall try to state clearly a~treatment of the fine-tuning based on the Wilsonian effective action and the corresponding Wilsonian renormalization group. Idea of the Wilsonian effective action is close to the intuitive understanding of the cut-off regularization. In a~standard discussion based on quadratic divergences, the artificial meaning of the scale of an~effective theory is given to the regularization parameter $\Lambda$. This effects in the regularization dependence of this kind of analysis. On the contrary, in the Wilsonian method high energy modes are integrated out in a~self-consistent, regularization-independent way, and an~effective theory has a~well-defined effective action. Moreover, this treatment is universal and depends very weakly on a~preferred UV completion. The main impact on the effective action from states with masses greater than the scale of the effective theory can be parametrized by the values of couplings of the Wilsonian effective action. Further corrections are highly suppressed as far as heavy masses are separated from the scale of the effective theory. 

Given a~model where the vacuum expectation value of a~scalar field can be generated with quantum corrections, we can also show how the stability of the effective action looks like from the point of view of the Wilsonian running. This has been compared with the standard picture based on the \mbox{Gell-Mann--Low} running. Since the Wilsonian running automatically includes integration of heavy degrees of freedom, the running differs markedly from the \mbox{Gell-Mann--Low} version. Nevertheless, similar behaviour can be observed: the scalar mass-squared parameter and the quartic coupling can change sign from a~positive to a~negative one due to running. This causes the spontaneous symmetry breaking or the instability in the renormalizable part of the potential for a~given range of scales. However, care must be taken when drawing conclusions, because of the truncation of higher dimension operators. The Gell-Mann--Low running allows one to resume relatively easily a class of operators corresponding to large logarithms to form the RGE improved effective potential valid over a huge range of scales. In the Wilsonian approach this would correspond to following the running of a large number of irrelevant operators, which is technically problematic.

While the simple cut-off analysis of scalar field models has been performed earlier, the goal of the present note is to consistently use the Wilsonian approach, and to make a~clear comparison with the discussion based on the Gell-Mann-Low running.
\section{Basic features of the model\label{model}}
\subsection{Couplings \label{Lagrangian}}
For the sake of clarity we consider a~simple model that exhibits certain interesting features of the SM. The model consists of a~massless Dirac fermion $\psi$ which couples via a~Yukawa interaction to a~real scalar field $\phi$ with a~quartic self-coupling. 
This Lagrangian takes the form:
\begin{equation}
\mathcal{L}=i \overline{\psi} \slashed{\partial} \psi + \frac{1}{2} \partial_\mu \phi \partial^\mu \phi - \frac{1}{2} M^2 \phi^2 - Y \phi \overline{\psi} \psi - \frac{\lambda}{4!} {\phi}^4. \label{Lagrangian_density}
\end{equation}
The above Lagrangian is symmetric under the (chiral) $\mathbb{Z}_2$ which acts on $\phi$ as $\phi\to-\phi$ and on $\psi$ as $\psi \to i \gamma^5 \psi$. We consider the case of the non-zero vacuum expectation value for the field $\phi$, which breaks this symmetry spontaneously. In the broken symmetry phase the Lagrangian density \eqref{Lagrangian_density} expanded around the nontrivial minimum $v$ ($\phi = v + \varphi$) will take the form:
\begin{equation}
\mathcal{L}=i \overline{\psi} \slashed{\partial} \psi - m \overline{\psi} \psi + \frac{1}{2} \partial_\mu \varphi \partial^\mu \varphi - \frac{1}{2} {M'}^2 \varphi^2 - Y \varphi \overline{\psi} \psi -\frac{g}{3!} \varphi^3 - \frac{\lambda}{4!} {\varphi}^4. \label{lagriangian_density_broken}
\end{equation}
The fermion $\psi$ allows one to model the top quark coupling to the Higgs boson, which is known to give the main contribution to quadratic divergencies in the mass of the SM scalar and to the high-scale instability of the quartic coupling. 

This model was previously investigated with methods of the FRG in \cite{Clark:1992jr,Clark:1994ya,Gies:2013fua}, in order to estimate the non-perturbative bound on the Higgs boson mass. The same issue was discussed in \cite{Gies:2014xha}, with a~slightly different Lagrangian. In \cite{Branchina:2005tu} the stability of potential was discussed with the help of the naive cut-off procedure. 

\subsection{Truncation adopted in the paper\label{truncation}}
We have calculated the Wilsonian renormalization group equations at the lowest non-trivial order. The Wilsonian action can include an infinite number of non-renormalizable operators, however they are suppressed at the low cut-off scale. Hence our truncation in the symmetric phase contains the following operators:
\begin{equation}
\mathcal{L}_{\Lambda}=i \overline{\psi}_{\Lambda} \slashed{\partial} \psi_{\Lambda} + \frac{1}{2} \partial_\mu \phi_{\Lambda} \partial^\mu \phi_{\Lambda} - \frac{1}{2} M^2_{\Lambda} {\phi_{\Lambda}}^2 - Y_{\Lambda} \phi_{\Lambda} \overline{\psi}_{\Lambda} \psi_{\Lambda} - \frac{\lambda_{\Lambda}}{4!} {\phi_{\Lambda}}^4. \label{truncation_Lagrangian}
\end{equation}
In the ordered phase it is convenient to use fluctuations $\varphi_{\Lambda}$ around the vacuum expectation value $v_{\Lambda}$. In such a case terms generated by expansion $\phi_{\Lambda} = v_{\Lambda} + \varphi_{\Lambda}$ must be included and our truncation takes the form:
\begin{equation}
\mathcal{L}_{\Lambda}=i \overline{\psi}_{\Lambda} \slashed{\partial} \psi_{\Lambda} - m_{\Lambda} \overline{\psi}_{\Lambda} \psi_{\Lambda} + \frac{1}{2} \partial_\mu \varphi_{\Lambda} \partial^\mu \varphi_{\Lambda} - \frac{1}{2} M^2_{\Lambda} {\varphi_{\Lambda}}^2 - Y_{\Lambda} \varphi_{\Lambda} \overline{\psi}_{\Lambda} \psi_{\Lambda} - \frac{g_{\Lambda}}{3!} {\varphi_{\Lambda}}^3 - \frac{\lambda_{\Lambda}}{4!} {\varphi_{\Lambda}}^4. \label{truncation_Lagrangian_ordered}
\end{equation}
Wilson coefficients $M^2_{\Lambda}$, $Y_{\Lambda}$, $\lambda_{\Lambda}$ as well as $m_{\Lambda}$ and $g_{\Lambda}$ are defined in \cite{Krajewski:2014vea}.

Chosen truncation corresponds to the lowest non-trivial order of the Wilsonian running (in the similar way as a~1-loop RGEs corresponds to lowest non-trivial order of the \mbox{Gell-Mann--Low} type running).

\section{Flow equations\label{RGE}}
It is convenient to express the Wilsonian RGE in terms of dimensionless parameters, because then Wilsonian RGEs are a~dynamical system of differential equations. We use the dimensionless parameters $\nu_{\Lambda}:=\frac{v_{\Lambda}}{{\Lambda}}$, ${\Omega_{\Lambda}}^2:=\frac{{M_{\Lambda}}^2}{{\Lambda}^2}$, $\omega_{\Lambda}:=\frac{m_{\Lambda}}{{\Lambda}}$ and $\gamma_\Lambda:= \frac{g_\Lambda}{{\Lambda}}$. We define $v_{\Lambda}$ by the requirement that the shift $\varphi_{\Lambda}\mapsto\phi_{\Lambda}-v_{\Lambda}$ gives the effective action without terms with odd powers\footnote{The standard treatment is such that in the ordered phase one expresses the symmetric Lagrangian density \eqref{truncation_Lagrangian} by the field shifted by its vev: $\phi \mapsto \varphi + v$. This shift introduces terms with the odd powers of $\varphi$ to the Lagrangian density. If we want to recover from the Lagrangian density expressed by $\varphi$, the value of vev by which $\phi$ was shifted, we need to search for the shift $\varphi \mapsto \phi - v$ such that terms with the odd powers of $\phi$ will be absent. Note that we expand around a~minimum, which means that the coefficient of the linear term vanishes.} of $\phi_{\Lambda}$.

Wilsonian $\beta$-functions for $\Omega^2_{\Lambda}$ and $\lambda_{\Lambda}$ in the ordered phase read as follows:
\begin{multline}
{\Lambda} \frac{d {\Omega_{\Lambda}}^2}{d{\Lambda}}=-2 {\Omega_{\Lambda}}^2+4\frac{Y_{\Lambda}^2}{(4 \pi) ^2} \left[\frac{\left(3-{\omega_{\Lambda}}^2\right) \left(1+4{\omega_{\Lambda}}^2\right) }{3 \left({\omega_{\Lambda}}^2+1\right)^3}{\Omega_{\Lambda}}^2-\frac{2{\omega_{\Lambda}}^2-2} {\left({\omega_{\Lambda}}^2+1\right)^2}\right]\\
+\frac{\lambda_{\Lambda}}{(4 \pi) ^2}\frac{1}{ ({\Omega_{\Lambda}}^2+1)}+\frac{{{\gamma}_{\Lambda}}^2}{(4 \pi) ^2}\left[\frac{2}{3}\frac{1}{1+{\Omega_{\Lambda}}^2}-\frac{2}{3}\frac{1}{(1+{\Omega_{\Lambda}}^2)^2}+\frac{{\Omega_{\Lambda}}^2}{(1+{\Omega_{\Lambda}}^2)^3}\right]\\
-{\gamma}_{\Lambda} \left[\frac{8Y_{\Lambda}}{(4 \pi )^2} \frac{\omega_{\Lambda}}{(1+{\omega_{\Lambda}}^2)}-\frac{{{\gamma}_{\Lambda}}}{(4 \pi )^2} \frac{1}{(1+{\Omega_{\Lambda}}^2)})\right],\label{RGE_M2}
\end{multline}
\begin{multline}
{\Lambda} \frac{d \lambda_{\Lambda}}{d{\Lambda}}=3 \frac{\lambda_{\Lambda}^2}{(4 \pi )^2}\frac{1}{(1+{\Omega_{\Lambda}}^2)^2}- \frac{2 Y_{\Lambda}^4}{(4 \pi )^2}\left[\frac{1}{\left(1+{\omega_{\Lambda}}^2\right)^2}-8\frac{{\omega_{\Lambda}}^2}{\left(1+{\omega_{\Lambda}}^2\right)^4}\right]+\frac{6{{\gamma}_{\Lambda}}^4}{(4 \pi )^2}\frac{1}{({\Omega_{\Lambda}}^2+1)^4}\\
+\frac{8}{3} \frac{\lambda_{\Lambda} Y_{\Lambda}^2 }{(4 \pi )^2} \frac{ \left(3-{\omega_{\Lambda}}^2\right)(1+4{\omega_{\Lambda}}^2)}{\left(1+{\omega_{\Lambda}}^2\right)^3}
+\frac{2{{\gamma}_{\Lambda}^2} \lambda_{\Lambda}}{(4 \pi )^2}\left[\frac{2}{3}\frac{1}{({\Omega_{\Lambda}}^2+1)^2}-\frac{7}{({\Omega_{\Lambda}}^2+1)^3}\right].\label{RGE_lambda}
\end{multline}
We obtained Wilsonian RGEs for discussed model using methods discussed in \cite{Kopietz:2010zz}.

\section{Numerical solutions of RGE\label{numerical}} 
The RGEs from Sec. \ref{RGE} have been solved numerically. We used the 1-loop matching conditions in order to compute initial conditions for RGE at $\Lambda=100$ (we use units of GeV through the paper), in terms of the measurable quantities (for definition see \cite{Krajewski:2014vea}).

\begin{figure}[!h]
\centering
\subfloat[]{\label{numerical_solution}
\includegraphics[width=0.5\textwidth]{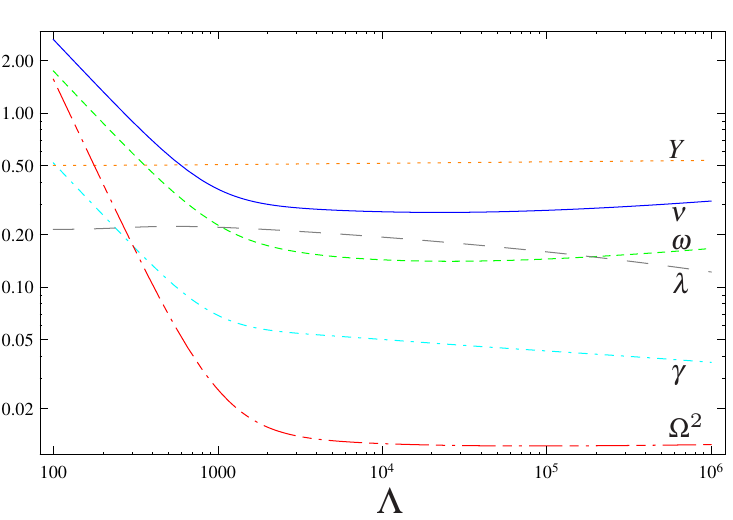}
}
\centering
\subfloat[]{\label{plot_Gell-Mann--Low}
\includegraphics[width=0.5\textwidth]{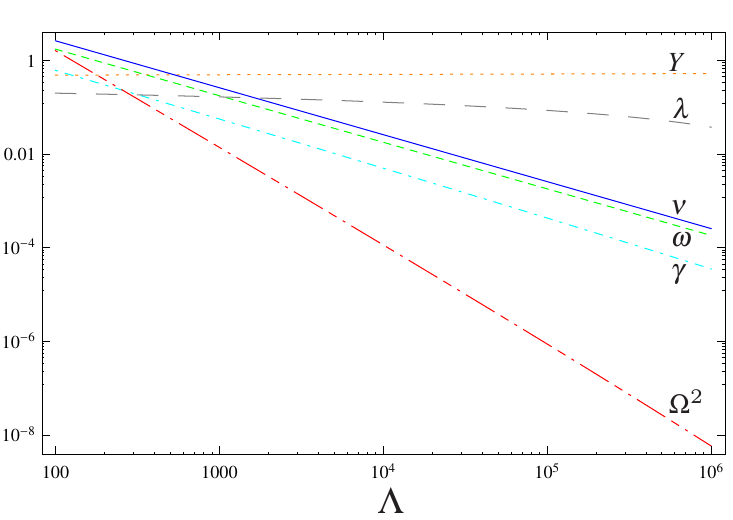}
}
\caption{Examples of a~numerical solution of \protect\subref{numerical_solution}---Wilsonian and \protect\subref{plot_Gell-Mann--Low}---\mbox{Gell-Mann--Low} type RGEs corresponding to: \mbox{$m_{ph}=174$}, \mbox{$M_{ph}=125$}, \mbox{$\lambda_{ph}=0.2$}, \mbox{$v_{ph}=264$}, \mbox{${g}_{ph}=52.8$} and \mbox{$Y_{ph}=0.5$}.}
\end{figure}

The example solution with the values: $m_{ph}=174$, $M_{ph}=125$, $\lambda_{ph}=0.2$, $v_{ph}=264$, ${g}_{ph}=52.8$ and $Y_{ph}=1$ is presented in Fig. \ref{numerical_solution}. Double-logarithmic plot in Fig. \ref{numerical_solution} shows parameters of the effective Wilsonian action as functions of the scale $\Lambda$. The dotted line represents the Yukawa coupling $Y_\Lambda$ which runs typically rather slowly. The quartic coupling $\lambda_{\Lambda}$ runs faster, because of the contribution from the fermionic loop. Couplings $\omega_\Lambda$, $\Omega^2_\Lambda$ and ${\gamma}_\Lambda$ for low values of $\Lambda$ run like relevant couplings due to the rescaling, but after reaching scales of the order of the masses, they change their behaviour to a~slow running near constant value. The behaviour of the above couplings for high values of $\Lambda$ is caused by quadratic divergences (or more precisely by the same diagrams which generate quadratic divergences). The same behaviour is manifested by the vacuum expectation value $\nu_\Lambda$ plotted as a~solid line.

\begin{figure}
 \begin{minipage}{.49\linewidth}
\centering
\subfloat[]{\label{numerical_omega}
\includegraphics[width=\textwidth]{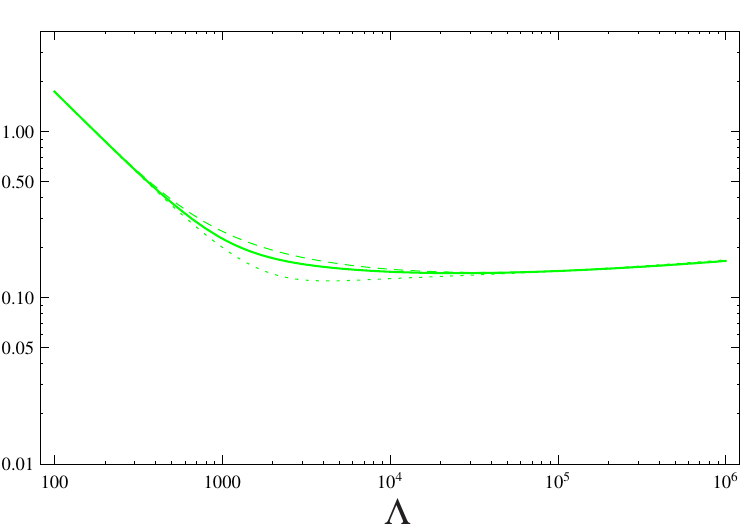}
 }
\end{minipage}
 \begin{minipage}{.49\linewidth}
\centering
\subfloat[]{\label{numerical_Omega}
\includegraphics[width=\textwidth]{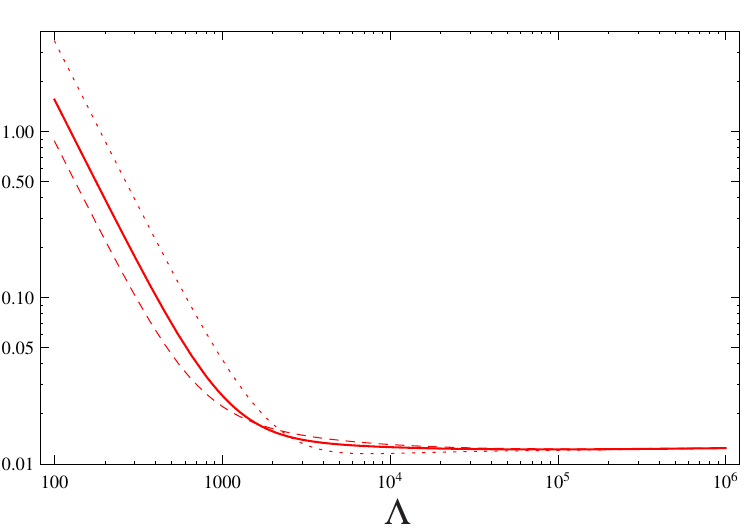}
 }
\end{minipage}
\par
 \begin{minipage}{.49\linewidth}
\centering
\subfloat[]{\label{numerical_g}
\includegraphics[width=\textwidth]{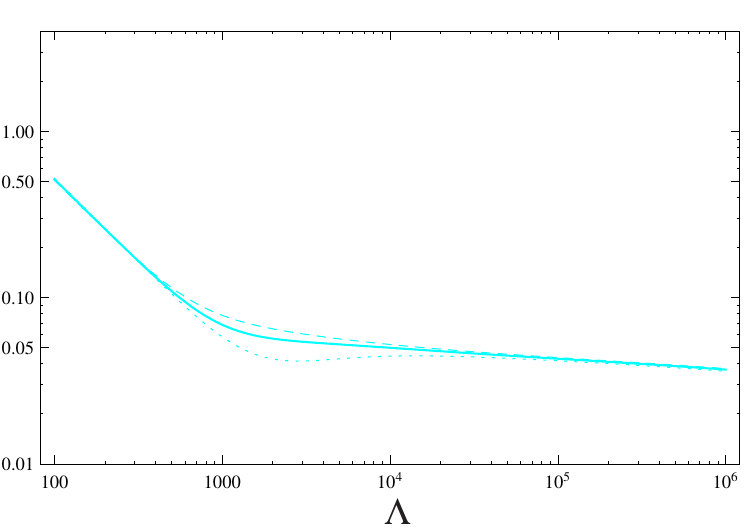}
 }
\end{minipage}
 \begin{minipage}{.49\linewidth}
\centering
\subfloat[]{\label{numerical_l}
\includegraphics[width=\textwidth]{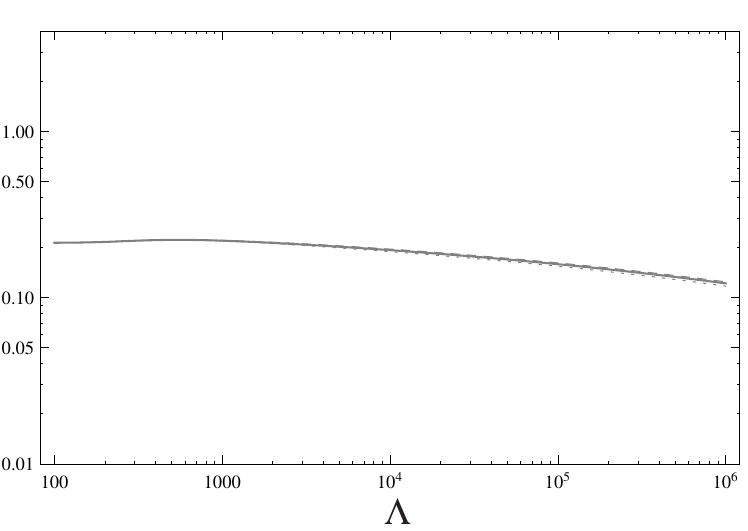}
 }
\end{minipage}
\caption{Solutions for the changed initial condition i.e. $M_{ph}$ multiplied (dashed) or divided (dotted) by the factor $\frac{3}{4}$, compared with the original one (solid line). The flow of $\omega_{\Lambda}$, $\Omega^2_{\Lambda}$, $\gamma_{\Lambda}$ and $\lambda_{\Lambda}$ is presented respectively in \protect\subref{numerical_omega}, \protect\subref{numerical_Omega}, \protect\subref{numerical_g} and \protect\subref{numerical_l}.}
\label{numerical_variation}
\end{figure}

Solutions with different initial conditions have the same qualitative behaviour. in Fig. \ref{numerical_variation} we plotted families of solutions with the initial conditions $M_{ph}=\frac{3}{4}125$ (dashed line) and $M_{ph}=\frac{4}{3}125$ (dotted line). The reference solution has been plotted as well with the solid lines.

The important observation is that the presented solutions run very close to each other for high scales $\Lambda$. Similar behaviour can be observed, if one changes the values of other parameters. Fig. \ref{numerical_variation} suggests the following understanding of the fine-tuning. If different trajectories of Wilsonian flow toward the UV approach each other, then the small change of parameters in the effective action at high scale, gives the effect of changing the trajectory. The change of values of the Wilson coefficients at high scale will result, in turn, in physical parameters much different than the original ones, since the IR limit of the Wilsonian running corresponds to measurable quantities. The behaviour presented in Fig. \ref{numerical_variation} is the sign that the fine-tuning of parameters in the effective action at high scales is required in order to get the prescribed values of physical observables.

The example of a~numerical solution for the \mbox{Gell-Mann--Low} type RGEs for the same theory is given in Fig. \ref{plot_Gell-Mann--Low}. Comparing Fig. \ref{plot_Gell-Mann--Low} with Fig. \ref{numerical_solution}, one finds that the flow of parameters of the Wilsonian effective action is much more complicated than the running in \mbox{Gell-Mann--Low} method. One should note that the Wilsonian RGE accommodates decoupling of massive particles i.e. corrections from particles with masses greater than $\Lambda$ are strongly suppressed. 

\section{Fine-tuning\label{fine_tuning}} 
The standard measure $\Delta_{c_i}$ of the fine-tuning with respect to the variable $c_i$ is defined as 
\begin{equation}
\Delta_{c_i} = \frac{\partial \log v^2}{\partial \log {c_i}^2},\label{finetuning_definition}
\end{equation}
where $c_i$ is a~coupling in the model and $v$ is the vacuum expectation value of the field which breaks a~symmetry spontaneously (here---chiral parity). 
As a~measure of the fine-tuning of the whole model we take \cite{Ellis:1986yg,Barbieri:1987fn,Ghilencea:2012gz}:
\begin{equation}
\Delta = \left( \sum_i {\Delta_{c_i}}^2\right)^{\frac{1}{2}}.\label{finetuning_definition_2}
 \end{equation} 
We have computed $\Delta_{c_i}$ for parameters of the effective action as functions of the scale $\Lambda$. 

Unfortunately, the effective action\footnote{Wilsonian effective action in the limit $\Lambda \to 0$ reproduce 1PI effective action.} for $\Lambda=0$ cannot be obtained by the direct numerical integration of RGEs.

For that reason we approximated $v_\Lambda$ for $\Lambda=0$ ($v_0$), by the value at $\Lambda=10^{-4}$ i.e. $v_{10^{-4}}$. We used $\Lambda=10^{-4}$, because this turns out to be the lowest scale which gives $\nu_\Lambda$ safety from numerical errors. 
To sum up, we have computed the fine-tuning measure \eqref{finetuning_definition} by taking numerical derivatives of $\nu_{10^{-4}}$ with respect to dimensionless parameters $\omega_\Lambda$, $\Omega^2_\Lambda$, ${\gamma}_\Lambda$, $\lambda_\Lambda$, $Y_\Lambda$ over the range of scales $10<\Lambda<10^6$. 

\begin{figure}
 \begin{minipage}[t]{.49\linewidth}
\centering
\includegraphics[width=\textwidth]{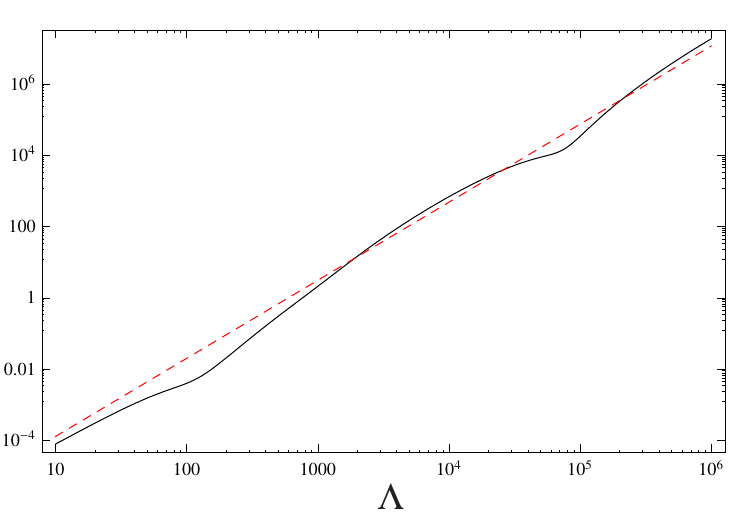}
\caption{ The combined fine-tuning measure \protect\eqref{finetuning_definition_2} as a~function of the scale $\Lambda$ of the Wilsonian effective action. Fitted power law is given as a~dashed line.\protect\label{finetuning_all}}
 \end{minipage}
 \begin{minipage}[t]{.49\linewidth}
 \centering
\includegraphics[width=\textwidth]{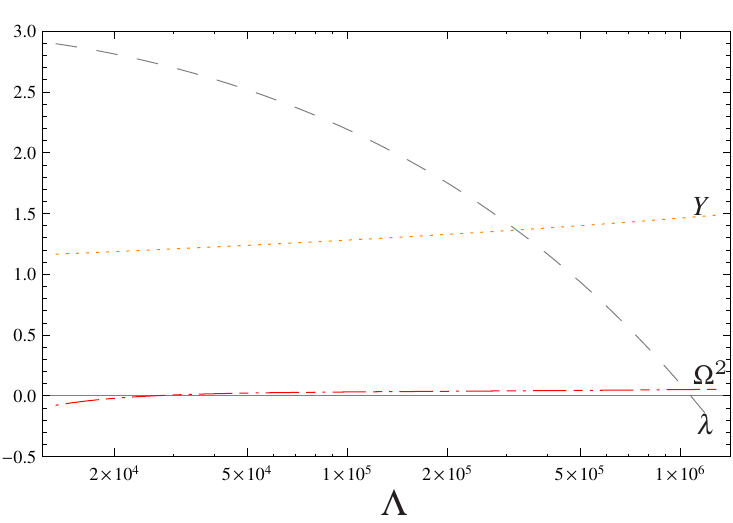}
\caption{The example of a~solution in which the radiative symmetry breaking takes place. The plot corresponds to the values $Y_{\Lambda}=1.461$, $M_{\Lambda}^2= 5 \times 10^{10}$, $\lambda_{\Lambda}=0.1$ at $\Lambda=10^6$.\protect\label{numerical_instability}}
\end{minipage}
\end{figure}

In Fig. \ref{finetuning_all} the measure \eqref{finetuning_definition_2} as a~function of scale $\Lambda$ is shown. The power function $\propto \Lambda^p$ which has been fitted to the fine-tuning curve is shown as a dashed line. The fitted power is equal $2.19\pm0.02$. The power-like function has been fitted over the interval $10^{3} \leq \Lambda \leq 10^{6} $ (that is above assumed mass thresholds). The reason is the visible change of behaviour of the flow of parameters below $10^{3}$.
\section{Vacuum stability\label{stability}} 
An interesting issue is the question of the spontaneous symmetry breaking and the stability of the potential seen from the point of view of the Wilsonian approach. 
The change of a~sign of $M_{\Lambda}^2$ during the flow toward IR indicates that the stable vacuum of the theory must have the non-zero vacuum expectation value of the scalar field $\phi$ (as long as the quartic coupling stays positive). Moreover, the quartic scalar coupling $\lambda_{\Lambda}$ can run negative for higher $\Lambda$ which shows the similar behaviour as the one observed in the \mbox{Gell-Mann--Low} type running (Fig. \ref{plot_Gell-Mann--Low}), known from the Standard Model. In the context of the SM, the zero of the quartic self-coupling is usually considered as an indication of the instability of the electroweak vacuum. 
In the Wilsonian approach however, simple analysis based on the quartic coupling alone is insufficient, because higher dimension operators with higher powers of the scalar field $\phi$, which we suppressed in our truncation \eqref{truncation_Lagrangian}, may dominate the scalar potential for large values of $\phi$. The impact coming from higher dimension operators was recently investigated in \cite{Gies:2013fua,Gies:2014xha,Eichhorn:2015kea} and \cite{Lalak:2014qua}.
To draw strong conclusions one needs a~procedure of resummation of the possibly large contributions to the scalar potential coming from operators with all higher powers of $\phi$. However, the observed instability of the Wilsonian quartic coupling may be seen as an indication of a~crossover behaviour at higher scales.

The example of a~solution demonstrating such features is plotted in Fig. \ref{numerical_instability}. For this solution the scalar mass parameter $\Omega_{\Lambda}^2$ vanishes at the scale $\Lambda=2.67 \times 10^4$ and the quartic coupling $\lambda_{\Lambda}$ has a~zero at $\Lambda = 1.07 \times 10^6$. While investigating features of this solution one can notice a~strong dependence of the scale of the symmetry breaking on the value of the Yukawa coupling $Y_{\Lambda}$. This fine-tuning problem makes one choose very precisely the initial condition for the Yukawa coupling in order to make the symmetry breaking scale low. 

\begin{figure}[!h]
\centering
\subfloat[The subspace of the phase space\protect\newline given by the \mbox{Eq. $Y=1.461$}.]{\label{phase_space_Omegavslambda}
\includegraphics[width=0.49\textwidth]{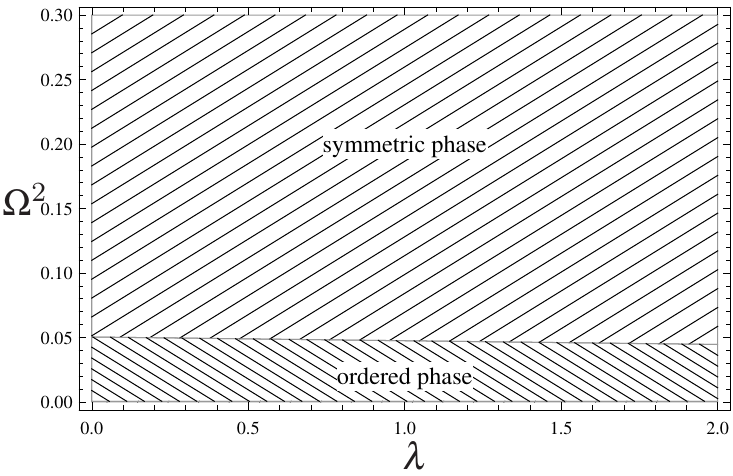}
}
\centering
\subfloat[The subspace of the phase space\protect\newline given by the \mbox{Eq. $\lambda=0.1$}.]{\label{phase_space_OmegavsY}
\includegraphics[width=0.49\textwidth]{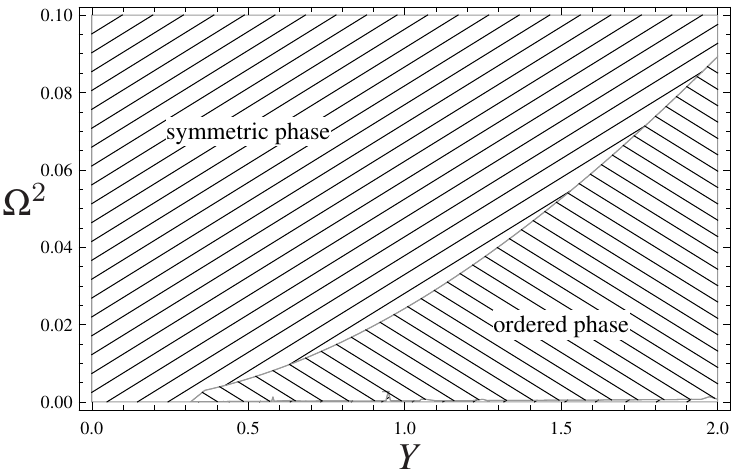}
}
\caption{The numerical approximation of the phase space of the model given by the Lagrangian density \protect\eqref{Lagrangian_density}. Points for which the flow from $\Lambda=10^{10}$ to $\Lambda=10^{-4}$ crosses $\Omega^2=0$ are assumed to be in the ordered phase.} \label{phase_space}
\end{figure}
The issue of the spontaneous symmetry breaking can be studied with the help of the numerical approximation of the phase space presented in Fig. \ref{phase_space}. Regions in Fig. \ref{phase_space} marked as "ordered phase" are points which used as initial conditions at $\Lambda=10^{10}$, produce the spontaneous breaking of chiral symmetry during the flow to $\Lambda=10^{-4}$. In Fig. \ref{phase_space_Omegavslambda} one can see that if for any scale $\Lambda$ the coupling $\Omega^2_{\Lambda}$ will be lower than certain critical value $\Omega^2_{cr}$, then $\Omega^2$ will run negative in the IR and the chiral parity will be spontaneously broken. Moreover as can be seen in \mbox{Fig. \ref{phase_space_OmegavsY}} the critical value $\Omega^2_{cr}$ is rather sensitive to the Yukawa coupling $Y$. From the behaviour shown in Fig. \ref{phase_space_OmegavsY} one concludes that $\Omega^2_{cr}$ increases when the value of $Y$ increases, and for any value of $\Omega^2$ there should exist a~critical value of the Yukawa coupling $Y_{cr}$. We conclude that once $Y_{cr}$ is exceeded, the radiative spontaneous symmetry breaking appears. Hence the lower right portion of Fig. \ref{phase_space_OmegavsY} gives the direct evidence of the Coleman--Weinberg mechanism at work. 

\section{Extended model\protect\label{decoupling}}
In order to father investigate the decoupling of heavy fields we considered an~extended model with two scalar fields $\phi_1$ and $\phi_2$, described by the Lagrangian density of the form:
\begin{equation}
\begin{split}
\mathcal{L}=& i \overline{\psi} \slashed{\partial} \psi + \frac{1}{2} \partial_\mu \phi_1 \partial^\mu \phi_1 + \frac{1}{2} \partial_\mu \phi_2 \partial^\mu \phi_2- \frac{1}{2} {M_1}^2 \phi_1^2 - \frac{1}{2} {M_2}^2 \phi_2^2\\
&- Y_1 \phi_1 \overline{\psi} \psi- Y_2 \phi_2 \overline{\psi} \psi\\
&- \frac{\lambda_1}{4!} {\phi_1}^4- \frac{\lambda_2}{4!} {\phi_2}^4- \frac{\lambda_3}{4} {\phi_1}^2 {\phi_2}^2- \frac{\lambda_4}{3!} {\phi_1} {\phi_2}^3- \frac{\lambda_5}{3!} {\phi_1}^3 {\phi_2}.
\end{split}
\label{Lagrangian_extended}
\end{equation}
In the limit $\lambda_3,\lambda_4,\lambda_5, Y_2 \to 0$ this theory reproduce previous theory given by the Lagrangian density \eqref{Lagrangian_density} if we identify the field $\phi_1 \mapsto \phi$ and couplings:
\begin{align}
M_1 &\mapsto M, & \lambda_1 &\mapsto \lambda, & Y_1 &\mapsto Y. \label{matching}
\end{align}
Hence one can supposes that for the heavy $\phi_2$ field (the case in which we are interested in), the Lagrangian density \eqref{Lagrangian_density} will describe effective theory for presented extended model.

The Lagrangian density \eqref{Lagrangian_extended} is symmetric under a~transformation $\phi_1 \to -\phi_1$, $\phi_2 \to -\phi_2$ and $\psi \to i \gamma^5 \phi$. Moreover if $Y_2=0$ ($Y_1=0$) simultaneously with $\lambda_4=\lambda_5=0$ a~transformation of $\phi_2$ ($\phi_1$) is independent from the transformation of remaining fields. Is such a case the Lagrangian density has two independent symmetries: $\phi_1 \to -\phi_1$ with $\psi \to i \gamma^5 \phi$ and $\phi_2 \to -\phi_2$ ($\phi_2 \to -\phi_2$ with $\psi \to i \gamma^5 \phi$ and $\phi_1 \to -\phi_1$). If any of $\lambda_4$, $\lambda_5$ or both Yukawa couplings $Y_1$, $Y_2$ are non-zero then fields $\phi_1$ and $\phi_2$ must have the same quantum numbers and can mix with each other. For the sake of clarity let us concentrate on this special case.

\subsection{Decoupling}
It is convenient to define the quantity:
\begin{equation}
{\rm nd}_i = \left|\frac{c_i^{EXT}-c_i^{EFF}}{c_i^{EXT}}\right| \label{difference}
\end{equation}
where the $c_i^{EXT}$ stands for one of the following couplings appearing in \eqref{Lagrangian_extended}: $m,M_1,\lambda_1,Y_1$. The $c_i^{EFF}$ is the coupling from the effective theory \eqref{Lagrangian_density} which corresponds to $c_i^{EXT}$ following the equation \eqref{matching}. The  quantities \ref{difference} are referred to as normalized differences.

\begin{figure}
\subfloat[]{\label{plotmatching_low}
\includegraphics[width=0.49\textwidth]{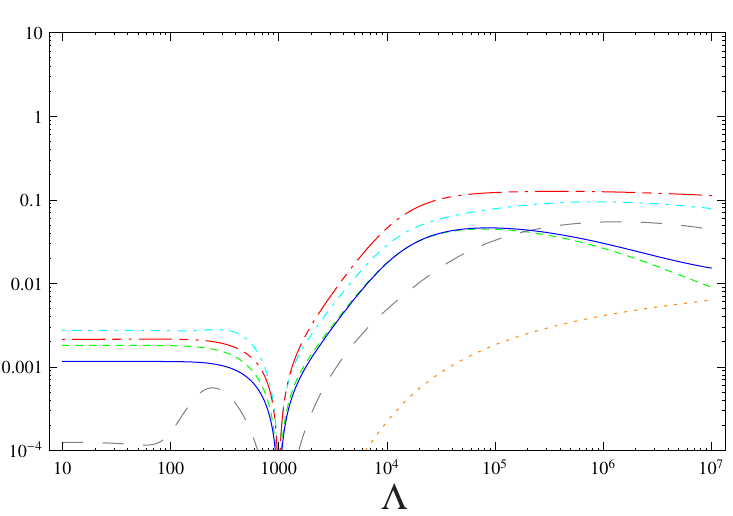}
}
\centering
\subfloat[]{\label{plotmatching_high}
\includegraphics[width=0.49\textwidth]{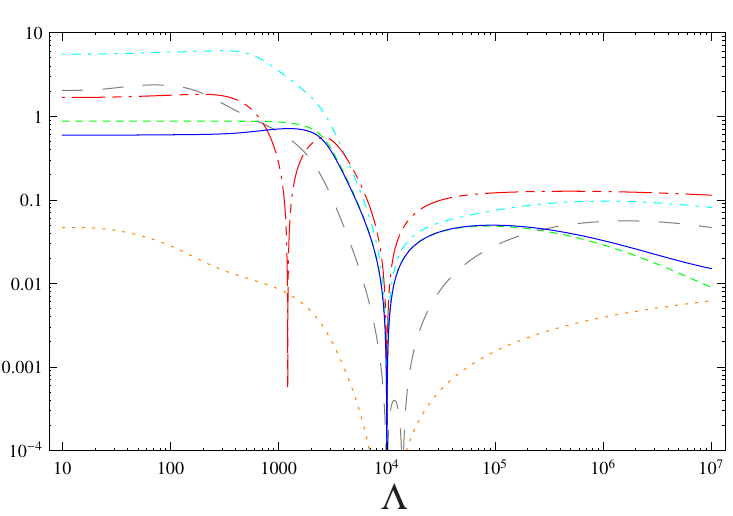}
}
\caption{Normalized difference \protect\eqref{difference} between running in effective theory and extended model for matching at two scales: \protect\subref{plotmatching_low}---$\Lambda_{matching}=M_2/10$ and \protect\subref{plotmatching_high}---$\Lambda_{matching}=M_2$.\label{difference_example}}
\end{figure}

Using Wilsonian RGEs for two theories, among which one is the effective theory for the other, one can try to find out how precise is the decoupling of heavy states in the running. We have estimated how Wilson coefficients change with scale in two scenarios. In the first case we use Wilsonian RGEs for the extended model to run from UV down to IR.  In the second case we use RGEs for extended model only down to an intermediate scale $\Lambda_{matching}$, lower then the mass $M_2$ of heavy scalar $\phi_2$. Below  $\Lambda_{matching}$ we use RGEs for the effective theory with initial conditions given by Wilson coefficients of extended model computed at the matching scale $\Lambda_{matching}$. In \mbox{Fig. \ref{difference_example}} quantities \eqref{difference} are plotted in two examples which differ by the choice of the scale $\Lambda_{matching}$. The spikes on both plots \ref{plotmatching_high} and \ref{plotmatching_low} correspond to the zeros i.e. the positions of matching scales $\Lambda_{matching}$ (because our matching condition is equality of couplings $c_i^{EXT}$ and $c_i^{EFF}$ at $\Lambda_{matching}$). In \mbox{Fig. \ref{plotmatching_high}} normalized differences as a functions of the scale $\Lambda$ in the effective action are plotted for the case of matching scale equal to the mass of the heavy scalar $\Lambda_{matching}=M_2=10^5$. In this case differences between Wilson coefficients in both theories in the IR limit are of the order of these couplings. In Fig. \ref{plotmatching_low} we plotted normalized differences for the matching scale $\Lambda_{matching}$ one decade lower (with the same initial conditions for RGEs in the extended model). One can notice that for the lower scale $\Lambda_{matching}$ quantities \ref{difference} fall down rapidly.
\section{Conclusions\label{conclusion}} 
In this paper we have used the Wilsonian effective action to investigate the fine-tuning and the vacuum stability in a~simple model exhibiting the spontaneous breaking of a~discrete symmetry and large fermionic radiative corrections which are able to destabilize the quartic scalar self-coupling. 
The regulator independence of the Wilsonian RG provides a~consistent and well-defined procedure to analyze the issue of quadratic divergences. In the simplified model simulating certain features of the SM, the Wilsonian renormalization group equations have been studied.

Numerical solutions of RGEs have revealed an interesting behaviour, caused by the same diagrams which generate quadratic divergences. An operator relevant near Gaussian fixed point (for example the mass parameter for the scalar particles) can run like a~marginal or even an~irrelevant operator, rather than decrease with growing scale. Furthermore solutions for different physical quantities flow close to each other with increasing scale. The flow in the direction of some common values indicates the severe fine-tuning. In such a~situation small changes of boundary values of parameters at high scale produce very different vacuum expectation values for the scalar field and other measurable quantities at low energies. We have estimated the fine-tuning as a~function of scale of the effective theory. For all parameters the adopted measure of the fine-tuning grows rapidly.

It should be stressed that the Wilsonian RGE, in contrast to the \mbox{Gell-Mann--Low} running, automatically accommodate decoupling of heavy particles. As investigated in Section \ref{decoupling}, the contribution to the flow coming from particles with masses $M_{heavy}$ greater than the scale $\Lambda$ of the effective action is strongly suppressed. The main contributions to the interactions generated by heavy states are integrated out during calculation of the effective action for $\Lambda \ll M_{heavy}$, and are included in the effective Wilson coefficients. 
These properties of Wilsonian RGE explain why the fine-tuning of the Wilson parameters is so interesting. Let us imagine a~more fundamental theory (say theory A) in which the SM is embedded. If one calculates in theory A the effective action for the scale $\Lambda$ below, but not very much below, the lowest mass of the particles from the New Physics sector, one obtains certain values of the Wilson coefficients $c^{A}$. On the other hand one can extrapolate the flow obtained from the SM to the scale $\Lambda$ and calculate the Wilson coefficients $c^{SM}$. Couplings computed in both ways should match, that is 
$c^{A}_{\Lambda} =c^{SM}_{\Lambda}$. 
If the couplings $c^A$ are different from $c^{SM}$ at the level of the fine-tuning $\Delta c$, that is $c^{A}_{\Lambda} (1 \pm \Delta c ) = c^{SM}_{\Lambda}$,
theory A will produce the IR effective action completely different from the \nolinebreak[4]SM.

We have studied the issue of the spontaneous symmetry breaking due to radiative corrections in the Wilsonian framework. We have demonstrated that there exists a~critical value $\Omega^2_{cr}$ below which $\Omega^2$ runs negative in the IR and the symmetry becomes spontaneously broken. Moreover, the critical value $\Omega^2_{cr}$ is sensitive to the Yukawa coupling $Y$. One can see that $\Omega^2_{cr}$ increases when the value of $Y$ increases and for any value of $\Omega^2$ there exists a~critical value of the Yukawa coupling $Y_{cr}$. Once $Y_{cr}$ is exceeded, the radiative spontaneous symmetry breaking appears, which is a~direct evidence of the Coleman--Weinberg mechanism at work. 

\acknowledgments{
This work has been supported by National Science Center under research grant DEC-2012/04/A/ST2/00099.
}
\bibliographystyle{JHEP}
\bibliography{FTVSWEA}

\providecommand{\href}[2]{#2}\begingroup\raggedright\begin{thebibliography}{10}

\bibitem{Aad:2012tfa}
{\bf ATLAS Collaboration} Collaboration, G.~Aad et~al., {\it {Observation of a
  new particle in the search for the Standard Model Higgs boson with the ATLAS
  detector at the LHC}},  {\em Phys.Lett.} {\bf B716} (2012) 1--29,
  [\href{http://arxiv.org/abs/1207.7214}{{\tt arXiv:1207.7214}}].

\bibitem{Chatrchyan:2012ufa}
{\bf CMS Collaboration} Collaboration, S.~Chatrchyan et~al., {\it {Observation
  of a new boson at a mass of 125 GeV with the CMS experiment at the LHC}},
  {\em Phys.Lett.} {\bf B716} (2012) 30--61,
  [\href{http://arxiv.org/abs/1207.7235}{{\tt arXiv:1207.7235}}].

\bibitem{Aoki:2012xs}
H.~Aoki and S.~Iso, {\it {Revisiting the Naturalness Problem -- Who is afraid
  of quadratic divergences? --}},  {\em Phys.Rev.} {\bf D86} (2012) 013001,
  [\href{http://arxiv.org/abs/1201.0857}{{\tt arXiv:1201.0857}}].

\bibitem{Farina:2013mla}
M.~Farina, D.~Pappadopulo, and A.~Strumia, {\it {A modified naturalness
  principle and its experimental tests}},  {\em JHEP} {\bf 1308} (2013) 022,
  [\href{http://arxiv.org/abs/1303.7244}{{\tt arXiv:1303.7244}}].

\bibitem{Jegerlehner:2013cta}
F.~Jegerlehner, {\it {The Standard model as a low-energy effective theory: what
  is triggering the Higgs mechanism?}},  {\em Acta Phys.Polon.} {\bf B45}
  (2014) 1167--1214, [\href{http://arxiv.org/abs/1304.7813}{{\tt
  arXiv:1304.7813}}].

\bibitem{deGouvea:2014xba}
A.~de~Gouvea, D.~Hernandez, and T.~M.~P. Tait, {\it {Criteria for Natural
  Hierarchies}},  {\em Phys.Rev.} {\bf D89} (2014) 115005,
  [\href{http://arxiv.org/abs/1402.2658}{{\tt arXiv:1402.2658}}].

\bibitem{Bar-Shalom:2014taa}
S.~Bar-Shalom, A.~Soni, and J.~Wudka, {\it {EFT naturalness: an effective field
  theory analysis of Higgs naturalness}},
  \href{http://arxiv.org/abs/1405.2924}{{\tt arXiv:1405.2924}}.

\bibitem{Clark:1992jr}
T.~E. Clark, B.~Haeri, S.~T. Love, M.~A. Walker, and W.~T.~A. ter Veldhuis,
  {\it {Wilson renormalization group analysis of theories with scalars and
  fermions}},  {\em Nucl.Phys.} {\bf B402} (1993) 628--656,
  [\href{http://arxiv.org/abs/hep-ph/9211261}{{\tt hep-ph/9211261}}].

\bibitem{Clark:1994ya}
T.~E. Clark, B.~Haeri, S.~T. Love, M.~A. Walker, and W.~T.~A. ter Veldhuis,
  {\it Mass bounds in the standard model},  {\em Phys. Rev. D} {\bf 50} (Jul,
  1994) 606--609.

\bibitem{Gies:2013fua}
H.~Gies, C.~Gneiting, and R.~Sondenheimer, {\it {Higgs Mass Bounds from
  Renormalization Flow for a simple Yukawa model}},  {\em Phys.Rev.} {\bf D89}
  (2014) 045012, [\href{http://arxiv.org/abs/1308.5075}{{\tt
  arXiv:1308.5075}}].

\bibitem{Gies:2014xha}
H.~Gies and R.~Sondenheimer, {\it {Higgs Mass Bounds from Renormalization Flow
  for a Higgs-top-bottom model}},  \href{http://arxiv.org/abs/1407.8124}{{\tt
  arXiv:1407.8124}}.

\bibitem{Branchina:2005tu}
V.~Branchina and H.~Faivre, {\it {Effective potential (in)stability and lower
  bounds on the scalar (Higgs) mass}},  {\em Phys.Rev.} {\bf D72} (2005)
  065017, [\href{http://arxiv.org/abs/hep-th/0503188}{{\tt hep-th/0503188}}].

\bibitem{Krajewski:2014vea}
T.~Krajewski and Z.~Lalak, {\it {Fine-tuning and vacuum stability in Wilsonian
  effective action}},  \href{http://arxiv.org/abs/1411.6435}{{\tt
  arXiv:1411.6435}}.

\bibitem{Kopietz:2010zz}
P.~Kopietz, L.~Bartosch, and F.~Schutz, {\it {Introduction to the functional
  renormalization group}},  {\em Lect.Notes Phys.} {\bf 798} (2010) 1--380.

\bibitem{Ellis:1986yg}
J.~R. Ellis, K.~Enqvist, D.~V. Nanopoulos, and F.~Zwirner, {\it {Observables in
  Low-Energy Superstring Models}},  {\em Mod.Phys.Lett.} {\bf A1} (1986) 57.

\bibitem{Barbieri:1987fn}
R.~Barbieri and G.~Giudice, {\it {Upper Bounds on Supersymmetric Particle
  Masses}},  {\em Nucl.Phys.} {\bf B306} (1988) 63.

\bibitem{Ghilencea:2012gz}
D.~M. Ghilencea, H.~M. Lee, and M.~Park, {\it {Tuning supersymmetric models at
  the LHC: A comparative analysis at two-loop level}},  {\em JHEP} {\bf 1207}
  (2012) 046, [\href{http://arxiv.org/abs/1203.0569}{{\tt arXiv:1203.0569}}].

\bibitem{Eichhorn:2015kea}
A.~Eichhorn, H.~Gies, J.~Jaeckel, T.~Plehn, M.~M. Scherer, and R.~Sondenheimer,
  {\it {The Higgs Mass and the Scale of New Physics}},  {\em JHEP} {\bf 04}
  (2015) 022, [\href{http://arxiv.org/abs/1501.02812}{{\tt arXiv:1501.02812}}].

\bibitem{Lalak:2014qua}
Z.~Lalak, M.~Lewicki, and P.~Olszewski, {\it {Higher-order scalar interactions
  and SM vacuum stability}},  \href{http://arxiv.org/abs/1402.3826}{{\tt
  arXiv:1402.3826}}.

\end{thebibliography}\endgroup
\end{document}